\documentclass{pasj01}

\begin{document}
\SetRunningHead{Itahana et al.}{Suzaku Observations of 1RXS J0603.3+4214}

\title{Suzaku Observations of the Galaxy Cluster 1RXS J0603.3+4214: Implications of Particle Acceleration Processes
       in ``Toothbrush'' Radio Relic}


%
 \author{
   Madoka \textsc{Itahana}\altaffilmark{1},
   Motokazu \textsc{Takizawa}\altaffilmark{2},
   Hiroki \textsc{Akamatsu}\altaffilmark{3},
   Takaya \textsc{Ohashi}\altaffilmark{4},
   Yoshitaka \textsc{Ishisaki}\altaffilmark{4},
   Hajime \textsc{Kawahara}\altaffilmark{5},
   Reinout J. \textsc{van Weeren}\altaffilmark{6}
   }
 \altaffiltext{1}{School of Science and Engineering, Yamagata University, 
                  Kojirakawa-machi 1-4-12, Yamagata 990-8560, Japan}
 \email{itahana@ksirius.kj.yamagata-u.ac.jp}
  \altaffiltext{2}{Department of Physics, Yamagata University, Kojirakawa-machi
                  1-4-12, Yamagata 990-8560, Japan}
 \email{takizawa@sci.kj.yamagata-u.ac.jp}
  \altaffiltext{3}{SRON Netherlands Institute for Space Research, Sorbonnelaan 2, 3584 CA Utrecht, The Netherlands}
  \altaffiltext{4}{Department of Physics, Tokyo Metropolitan University, 1-1 Minami-Osawa, Hachioji, Tokyo 192-0397, Japan}
  \altaffiltext{5}{Department of Earth and Planetary Science, The University of Tokyo, 7-3-1 Hongo, Bunkyo-ku, Tokyo 133-0033, Japan}
  \altaffiltext{6}{Harvard-Smithsonian Center for Astrophysics, 60 Garden Street, Cambridge, MA 02138, USA}


\maketitle

\begin{abstract}
We present the results of Suzaku observations of the galaxy cluster 1RXS J0603.3+4214 with ``toothbrush'' radio relic.
Although a shock with Mach number $M \simeq 4$ is expected at the outer edge of the relic from the radio observation,
our temperature measurements of the intracluster medium indicate a weaker temperature difference than what is expected.
The Mach number estimated from the temperature difference at the outer edge of the relic is $M \simeq 1.5$, which is significantly
lower than the value estimated from the radio data even considering both statistical and systematic errors.
This suggests that a diffusive shock acceleration theory in the linear test particle regime, which is commonly used to link
the radio spectral index to the Mach number, is invalid for this relic. 
We also measured the temperature difference across the western part of the relic, where a shock with $M \simeq 1.6$ is suggested 
from the X-ray surface brightness analysis of the XMM-Newton data, and obtained consistent results in an independent way.
We searched for the non-thermal inverse Compton component in the relic region and the resultant upper limit on the flux
is $2.4 \times 10^{-13}$ erg cm$^{-2}$ s$^{-1}$ in the 0.3-10 keV band. The lower limit of the magnetic field
strength becomes 1.6 $\mu$G, which means that magnetic energy density could be more than a few $\% $ of the thermal energy.
\end{abstract}

\section{Introduction}
The standard theory of cosmological structure formation tells us that rich galaxy clusters form through
mergers and absorption of smaller galaxy clusters and groups. Cluster major mergers are the largest energy release
events in the universe where the energy of $\sim 10^{64}$ erg is involved. Mergers have great impacts on cluster evolution.
Indeed, numerical simulations \citep{Rick01,Taki05,Akah10,Taki10} show that mergers cause shocks and turbulence 
in the intracluster medium (ICM). 
While a large part of the dissipated energy in the shocks and turbulence is transformed into the thermal ICM, 
some are most likely transported into a non-thermal form such as cosmic-rays and magnetic fields 
\citep{Ohno02,Taki08,Zuho11,Donn13}.

In fact, some merging clusters have diffuse non-thermal synchrotron radio emissions, which is direct evidence
of the existence of the cosmic-ray electrons whose energy is $\sim$ GeV and the magnetic fields of $\sim \mu$G 
in the ICM. Such diffuse non-thermal radio sources are classified into two types owing to their
morphology and location. Radio halos have similar morphology as the ICM X-ray emission and are located in the central part
of the cluster \citep{Fere97,Lian00,Govo04,vanW11}. 
On the other hand, radio relics usually have an arc-like shape and are located in the
periphery of the cluster \citep{Rott97,Bona09,vanW10,van Weeren2012}. 
The recent status of this field is extensively reviewed in \citet{Fere12} and \citet{Brun14}.

Because of their morphology and locations, it has been believed that radio relics trace outgoing shocks 
seen in the later phases of major mergers. Recently, in fact, discontinuities of the ICM physical quantities such as temperature 
and density are found in the outer edge of some radio relics 
\citep{Fino10,Maca11,Akam12a,Akam12b,A&K13,Akam13,ogrean2013,Shim15},
which is direct evidence of the association of relics with shocks.
Mach numbers of shocks related to the relics can be observationally estimated in two independent ways.
From the synchrotron radio spectral index, we are able to obtain the cosmic-ray electron energy spectral index \citep{Rybi79},
which can be related with the shock Mach numbers assuming a simple diffusive shock acceleration (DSA) theory \citep{Drur83,Blan87}.
On the other hand, X-ray observations of the ICM are able to determine directly the Mach number through the detection of
discontinuities of the ICM physical quantities such as temperature and density with Ranking-Hugoniot jump conditions \citep{Land59,Shu92}.
If the simple diffusive shock acceleration model is valid, both methods have to give us consistent results. 
Thus, comparison of the high quality X-ray data with excellent radio spectra is useful to get insights into 
the particle acceleration processes. \citet{A&K13} performed the first systematic study about this issue, 
though the sample size was not large enough to draw definite conclusions. 
Therefore, it is important to observe more relics in both radio and X-rays in order to reveal their origin.

The electron population attributed to the synchrotron radiation also emits non-thermal X-rays through
the inverse Compton processes of cosmic microwave background (CMB) photons. 
Comparison between synchrotron and inverse Compton fluxes provides us with the magnetic field strength. 
Even if only the upper limit of the inverse Compton flux is obtained, we are able to get the lower limit of the field strength.
Although a lot of efforts have been made to search for
such non-thermal X-rays from cluster radio halos and relics \citep{Ajel09,Kawa09,Naka09,Suga09,Ajel10,Wik11}, 
this is still challenging and any firm detections have not been reported unfortunately.

1RXS J0603.3+4214 is a merging cluster with radio relics and a halo at $z=0.225$. 
Its north radio relic is well-known for a peculiar 
linear shape and nicknamed by ``toothbrush'' \citep{van Weeren2012}. The radio spectrum at the outer edge of the relic is rather flat
and its spectral index is $\alpha = -0.6 \sim -0.7$, which means that the shock Mach number is $M = 3.3 \sim 4.6$.
This value is rather high for cluster merger shocks. An XMM-Newton observation reveals that it has an elongated north-south 
morphology with two subcluster cores \citep{ogrean2013}, which strongly suggests that this cluster is in the later phase of 
a major merger. 
From the X-ray surface brightness distribution analysis, it is claimed that the shock is offsetted 
by 1 arc-minute outwards from the outer edge of the relic and extended in the west of the relic. The Mach number is estimated 
to be $\sim 1.7$ from the density modeling with a discontinuity,
which is significantly lower than the value derived from the radio data. However, note that their results could be affected 
by the three dimensional density structure modeling and projection effects.

We conducted a Suzaku X-ray observation of the field around the ``toothbrush'' relic. The X-ray Imaging Spectrometer (XIS)
aboard the Suzaku satellite \citep{Mits07} is more appropriate for observing low surface brightness diffuse sources 
than Chandra and XMM-Newton, thanks to its low and stable background \citep{Koya07}. 
This feature is very useful to investigate the physical properties of the ICM 
around cluster radio relics, which are generally located in the outerpart of the cluster. 
Indeed, ICM temperature in the pre-shock region around radio relics can be determined practically only by Suzaku,
considering that the ICM emission fill the field of view and in the hard band.

In this paper, we present a Suzaku X-ray observation of galaxy cluster 1RXS J0603.3+4214 to investigate the physical status and 
particle acceleration processes at ``toothbrush'' radio relic. The rest of this paper is organized as follows. 
In section 2 we describe the observation and data reduction. In section 3 we present spectral analysis results. 
In section 4 we discuss the results and their implications. In section 5 we summarize the results. 
Canonical cosmological parameters of $H_0 = 70$ Mpc$^{-1}$ km s$^{-1}$ , $\Omega_0 = 0.27$, and $\Lambda_0 = 0.73$ are used 
in this paper. At the redshift of the cluster, 1 arcmin corresponds to 217 kpc.
The solar abundances are normalized to \citet{Ande89}.
Unless otherwise stated, all uncertainties are given at the 90\% confidence level. 

\section{Observations and Data Reduction}

We observed a field around ``toothbrush'' radio relic in galaxy cluster 1RXS J0603.3+4214 with Suzaku on 2012 October 7-10.
In addition, a field offsetted by 1 degree in Galactic longitude was observed on 2012 October 14-15 to estimate the background components. 
The observations were performed at XIS nominal pointing. The XIS was operated in the normal full-frame clocking mode. 
The edit mode of the data format was $3 \times 3$ and $5 \times 5$, and combined data of both modes were utilized. 
The spaced-row charge injection was adopted for XIS. 
All data were processed with Suzaku pipeline processing, version 2.8.16.34. The calibration data files (20130305) were adopted.
A summary of the observations is given in table \ref{tab:obs}.
\begin{table*}
  \caption{An observational log of 1RXS J0603.3+4214}\label{tab:obs}
  \begin{center}
    \begin{tabular}{cccc} \hline
      Name (Obs.ID)                 & (RA, Dec)        & Observation Date & Exposure (ksec)\\ \hline
      1RXS J0603.3+4214 (807001010) & (\timeform{90D.7885}, \timeform{+42D.2628}) & 2012/10/7-10     & 124.4 \\
      OFFSET (807002010)            & (\timeform{92D.2826}, \timeform{+42D.2613}) & 2012/10/14-15    & 26.8 \\ \hline 
    \end{tabular}
  \end{center}
\end{table*}

The XIS data were processed through standard criteria as follows. Events with a GRADE of 0, 2, 3, 4, 6 and STATUS with 0:524287 were 
extracted. Data obtained at the South Atlantic Anomaly (SAA), within 436 s after the passage of SAA, 
and at low elevation angles from an Earth rim of $< 5^{\circ}$ and a Sun-lit Earth rim of $< 20^{\circ}$ were excluded. The resultant
effective exposure times were 124.4 ks and 26.8 ks for the cluster and offset region, respectively.
We applied additional processing for XIS1 to reduce the non X-ray background (NXB) level, which increased after changing 
the amount of charge injection, following descriptions in the XIS analysis topics 
(see \verb|http://www.astro.isas.jaxa.jp/suzaku/analysis/xis/| \\ \verb|xis1_ci_6_nxb/|).
NXB spectra and images of XIS were generated using the ftool ``xisnxbgen'' \citep{Tawa08}. 
We tried to perform event screening with cut-off rigidity to improve the signal-to-noise ratio. 
However, the decrease of the NXB systematic errors did not compensate the increase of statistical errors in this procedure.

\begin{figure}
  \begin{center}
    \includegraphics[width=8cm]{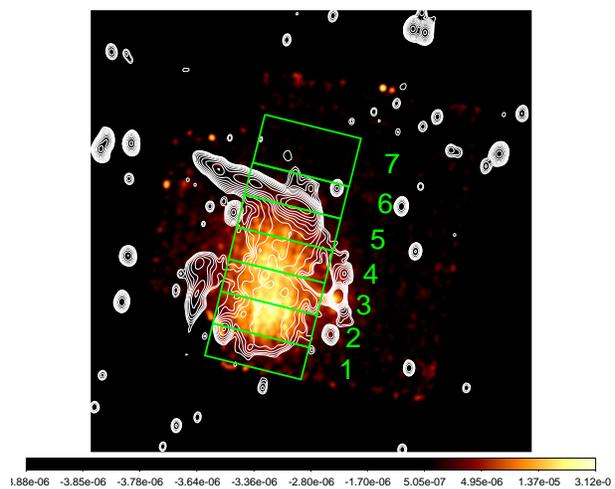} 
  \end{center}
  \caption{An XIS3 image of 1RXS J0603.3+4214 cluster in the 0.5-8.0 keV band overlaid with the 1.16-1.78 GHz radio contours
           \citep{van Weeren2012}. The X-ray image was corrected for exposure and vignetting effects after subtracting NXB, 
            and smoothed it by a Gaussian kernel with $\sigma=0.26'$. The regions used 
           to investigate temperature distribution along the collision axis in subsection \ref{ss:tempcolaxis} are displayed
           as green boxes. The region numbers are also shown in green.}
           \label{fig:Toothbrush}
\end{figure}

For all XIS CCD chips, fields corresponding to the damaged part of XIS0 CCD chip because of a micrometeorite accident were not utilized 
in the following analysis. 
Figure \ref{fig:Toothbrush} represents an 0.5-8.0 keV XIS3 image overlaid with the 1.16-1.78 GHz radio contours 
\citep{van Weeren2012}. The X-ray image was corrected for exposure and vignetting effects after subtracting NXB, 
and smoothed it by a Gaussian kernel with $\sigma=0.26'$.

\section{Spectral Analysis}

For the spectral analysis of the XIS data, redistribution matrix files (RMFs) were generated using the ftool ``xisrmfgen''. 
In addition, ancillary response files (ARFs), which describe the response of X-ray telescope aboard Suzaku and 
the amount of the XIS optical blocking filters contamination, were generated with the ftool ``xissimarfgen'' \citep{Ishi07}. 
Uniform emission over a circular region with $20'$ radius was used as an input image to generate an ARF. 
For the spectral fitting of the background and cluster fields, we used the energy bands of 0.5-8.0 keV and 0.7-7.0 keV,
respectively, ignoring the energy range near the Si edge (1.7-1.8 keV) to avoid uncalibrated structures.
The XIS spectra for each sensor (XIS0, XIS1, XIS3) were fitted simultaneously.

\subsection{The Background}\label{ss:BGD}
In order to perform the spectral analysis of the cluster, a background model from a celestial origin have to be estimated in addition to the NXB. 
Because all of the field of view (FOV) of the cluster is expected to be filled by the ICM emission,
We utilized the 1 degree offset region for this purpose. The cosmic X-ray background (CXB), the Milky Way halo's hot gas (MWH), 
and the local hot bubble (LHB) were taken into account as the background components.
Then, the spectrum of the background region was fitted using the following model:
\begin{eqnarray}
  apec_{\rm LHB}+wabs*(apec_{\rm MWH}+powerlaw_{\rm CXB}),
\end{eqnarray}
where $apec_{\rm LHB}$, $apec_{\rm MWH}$, and $powerlaw_{\rm CXB}$ represent the LHB, MWH, and CXB, respectively. 
The temperature of the LHB was fixed to be 0.08 keV. The metal abundance and redshift of both LHB and MWH were also fixed to be solar
and zero, respectively. The powerlaw index of the CXB was fixed to be 1.4 \citep{Kush02}. 
We assumed $N_{\rm H} = 1.79 \times 10^{21}$ cm$^{-2}$ for the Galactic absorption \citep{Dick90}. 

\begin{figure}
  \begin{center}
    \includegraphics[angle=270,width=8cm]{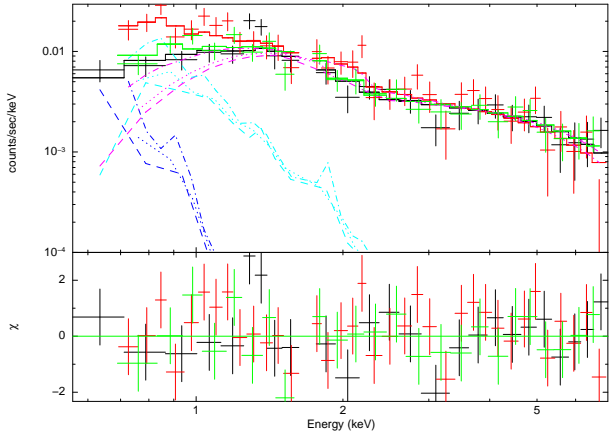}
  \end{center}
  \caption{The XIS spectra of the background field fitted with the background model described in the text.
         The black, red, and green crosses show the spectra of XIS0, XIS1, and XIS3, respectively.
         The total spectra of the best-fit model are also plotted as solid histograms.
         The blue, light blue, and magenta dashed histograms represent the LHB, MWH, and CXB components, respectively.}
         \label{fig:BGD}
\end{figure}

The spectra of the background field fitted with the above-mentioned model are shown in figure \ref{fig:BGD}, 
where black, red, and green crosses show the spectra of XIS0, XIS1, and XIS3, respectively.
The total and each component of the best-fit model spectra are also plotted as solid and dashed histograms, respectively.
The detailed results of the fit are summarized in table \ref{tab:BGD}. 
The obtained temperature of the MWH component ($kT=0.56^{+0.12}_{-0.21}$ keV)
is somewhat higher than the typical value ($kT \sim 0.3$ keV), which might be due to a patchy high temperature component
found in the former analysis \citep{Yosh09,Seki14}.

\begin{table*}
  \caption{Best-fit parameters for the XIS spectra of the background field}\label{tab:BGD}
  \begin{center}
    \begin{tabular}{ccc} \hline 
      Model Component   & Parameter                          & Value
     \\ \hline
      LHB               & $kT$\footnotemark[$*$]             & 0.08 (fixed) \\
                        & $N$\footnotemark[$\dagger$]        & $5.91^{+3.56}_{-3.78} \times 10^{-2}$   \\ \hline
      MWH               & $kT$\footnotemark[$*$]             & $0.56^{+0.12}_{-0.21}$           \\
                        & $N$\footnotemark[$\dagger$]        & $3.94^{+3.02}_{-1.03} \times 10^{-4}$    \\ \hline
      CXB               & $\Gamma$\footnotemark[$\ddagger$]  & 1.4 (fixed)  \\
                        & $N$\footnotemark[$\S$]             & $9.05^{+0.59}_{-0.58} \times 10^{-4}$     \\ \hline  
                        & $\chi^2/d.o.f$                     & $72.67/81$     \\ \hline 
    \\
   \multicolumn{2}{@{}l@{}}{\hbox to 0pt{\parbox{180mm}{\footnotesize
       \footnotemark[$*$] Temperature of the each component in keV.
       \par\noindent
       \footnotemark[$\dagger$] Normalization in the $apec$ code for each component scaled with a factor $1/400 \pi$. \\
                     $N=(1/400 \pi) \int n_{\rm e} n_{\rm H} dV / [ 4 \pi (1+z)^2 D_{\rm A}^2 ] \times 10^{-14}$ cm$^{-5}$ arcmin$^{-2}$,\\
                                where $D_{\rm A}$ is the angular diameter distance to the source.
       \par\noindent
       \footnotemark[$\ddagger$] Photon index of the power-law component.
       \par\noindent
       \footnotemark[$\S$] Normalization in the power-law component in photons keV$^{-1}$ cm$^{-2}$ s$^{-1}$ at 1 keV
     }\hss}}
    \end{tabular}
  \end{center}
\end{table*}

\subsection{Temperature Distribution along the Collision Axis}\label{ss:tempcolaxis}
\begin{figure}
  \begin{center}
    \includegraphics[angle=270,width=8cm]{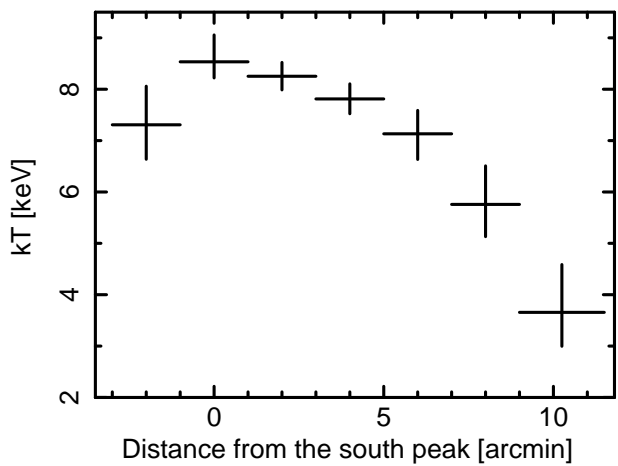}
  \end{center}
  \caption{
      The temperature profile along the collision axis. The horizontal axis shows the angular distance from the
      southern main X-ray peak (region 2 in figure \ref{fig:Toothbrush} with the highest temperature), where the north
      and south directions are positive and negative, respectively
  }
  \label{fig:temp_plot}
\end{figure}

Before performing detailed analysis of the temperature structure around candidate shock regions,
we derive the temperature distribution along the collision axis to reveal the overall structure of the system.
The regions used for this purpose are presented in figure \ref{fig:Toothbrush}. The spectrum for each region is fitted 
by the following model;
\begin{eqnarray}
 const*[apec_{\rm LHB}+wabs*(apec_{\rm MWH}+powerlaw_{\rm CXB}\\ \nonumber +apec_{\rm ICM})],
\end{eqnarray}
where, $apec_{\rm ICM}$ represents the emission from the ICM. All the parameters in the background components 
($apec_{\rm LHB}$, $apec_{\rm MWH}$, and $powerlaw_{\rm CXB}$) are fixed to the values derived from the background field
in subsection \ref{ss:BGD}. 
We assumed that $N_{\rm H} = 2.14 \times 10^{21}$ cm$^{-2}$ for the Galactic absorption \citep{Dick90}
and $z=0.225$ for the redshift of the ICM. We fixed the metal abundance to be 0.2 solar for regions 
where we did not have enough photons to determine it. 
We introduce $const$ to take into account of possible calibration uncertainty among different XIS sensors, 
which is fixed to be unity for XIS1 and allowed to vary freely for XIS0 and XIS3. The other parameters except $const$ are common among 
all XIS sensors. The most results of $const$ for XIS0 and XIS3 differ only by less than $10\%$ from unity, which is consistent with 
the expected calibration uncertainty, except for a few cases. We also fit the spectra by a model without $const$ and compared both results 
with each other. Although the resultant chi squares are different, numerical values of the best fit parameters are consistent with each other 
within statistical errors. 
The fitting results are presented in table \ref{tab:temp}.
The temperature is the highest in the region 2, where the southern main X-ray peak is located.
The obtained temperature profile along the collision axis is shown in figure \ref{fig:temp_plot}, 
where the horizontal axis shows the angular distance from the southern main X-ray peak 
(region 2 in figure \ref{fig:Toothbrush} with the highest temperature).
The north and south directions are positive and negative, respectively.
The temperature seems to decline smoothly with the distance from the peak.

\begin{table}
  \caption{Best fit parameters for the XIS spectra of the region presented in figure \ref{fig:Toothbrush}}\label{tab:temp}
  \begin{center}
  \begin{tabular}{cccc} \hline
  Region Number & $kT$ (keV)         & abundance           & $\chi^2/d.o.f$ \\ \hline
    1           &$7.31^{+0.75}_{-0.67}$ & $0.24^{+0.08}_{-0.08}$ & $174.37 /189 $  \\
    2           &$8.53^{+0.52}_{-0.31}$ & $0.18^{+0.03}_{-0.03}$ & $819.42 /739 $            \\ 
    3           &$8.25^{+0.27}_{-0.26}$ & $0.16^{+0.03}_{-0.03}$ & $1043.07 /925 $            \\
    4           &$7.81^{+0.29}_{-0.29}$ & $0.13^{+0.03}_{-0.03}$ & $908.03 /814 $            \\ 
    5           &$7.13^{+0.46}_{-0.50}$ & $0.16^{+0.04}_{-0.04}$ & $461.81 /398 $   \\
    6           &$5.76^{+0.75}_{-0.63}$ & $0.22^{+0.11}_{-0.10}$ & $116.45 /126 $            \\ 
    7           &$3.66^{+0.93}_{-0.66}$ & $0.2 ({\rm fixed})$   & $67.93 /80 $      \\  \hline
  \end{tabular}
  \end{center}
\end{table}

\subsection{Analysis of the Candidate Shock Regions}\label{ss:an_ca_sh_re}
\begin{figure}
  \begin{center}
    \includegraphics[width=8cm]{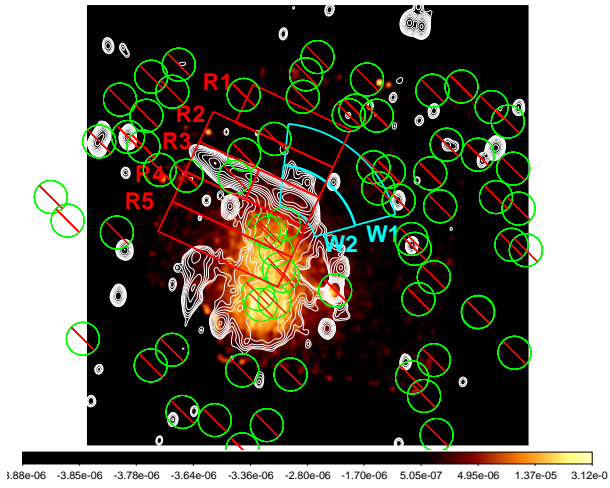}
  \end{center}
  \caption{Regions utilized in the candidate shock analysis. Red boxes and cyan annuli are regions for the analysis 
    of the relic shock and west shock, respectively. Green circles are excluded regions of a point source.}
  \label{fig:region}
\end{figure}

In this subsection, we investigate temperature structure around the candidate shock regions in more detail.
The spectral model used in the fitting is the same as in subsection \ref{ss:tempcolaxis}. 
Figure \ref{fig:region} shows regions used in the analysis. 
We searched for point sources with XMM-Newton data, whose spatial resolution is better than Suzaku. To reduce
contamination and CXB systematic errors, $1'$ radius circular regions centered by a position of a point source whose flux is more than 
$1.0 \times 10^{-14}$ erg cm$^{-2}$ s$^{-1}$ are excluded and shown in figure \ref{fig:region} by green circles.
It is expected that an outgoing shock front exits at the outer edge of the radio relic. 
We chose the regions as $2' \times 8'$ boxes, shown in red in figure \ref{fig:region},
except for the outermost region, where a $2.5' \times 6.5'$ box is adopted to avoid the region illuminated by the calibration sources, 
in order to investigate temperature profile across the expected shock, assuming that the shock is located 
at the relic's outer edge. 
A temperature gradient along the relic's long axis is suggested by \citet{ogrean2013}. 
Region R3 is divided into two $2' \times 4'$ boxes as 
in figure \ref{fig:region} to investigate how this affects the Mach number estimation.
\citet{ogrean2013} also suggests that the shock is elongated towards the west of the relic.
Cyan annuli in figure \ref{fig:region} are regions to investigate this expected west shock.

Systematic errors of both CXB and NXB were estimated in a similar way in \citet{Wata11}. The CXB spatial fluctuations 
can be modeled as $\sigma_{\rm CXB}/I_{\rm CXB} \propto \Omega_{\rm e}^{-0.5} S_{\rm c}^{0.25}$,
where $\Omega_{\rm e}$ and $S_{\rm c}$ are the effective solid angle and upper cutoff flux of a point source, respectively.
From the HEAO-I A2 results,  \citet{Shaf83} reported that
\begin{eqnarray}
  \frac{\sigma_{\rm CXB}}{I_{\rm CXB}} = 2.8 \% \biggl( \frac{\Omega_{\rm e}}{15.8 \ {\rm deg}^2} \biggr)^{-0.5} 
                                     \biggl( \frac{S_{\rm c}}{8 \times 10^{-11} \ {\rm erg \ s}^{-1} {\rm cm}^{-2} } \biggr)^{0.25}.
  \label{eq:CXB}
\end{eqnarray}
Adopting an upper cutoff flux of $S_{\rm c} = 1.0 \times 10^{-14}$ erg cm$^{-2}$ s$^{-1}$ and the values of the regions used 
in the spectral fit for $\Omega_{\rm e}$, we calculated the CXB fluctuations at the 90 \% confidence level,
which are shown in table \ref{tab:relic} and \ref{tab:west} for each region.
On the other hand, it is reported that the reproducibility of NXB was 6.0 \% and 12.5 \% for XIS FI and BI at the 90 \% confidence 
level, respectively \citep{Tawa08}. Systematic errors of LHB and MWH were not considered because the ICM temperature is sufficiently higher
than those of the Galactic background components and it is most likely that the effects of those errors are very limited.

\begin{figure}
  \begin{center}
    \includegraphics[angle=270,width=8cm]{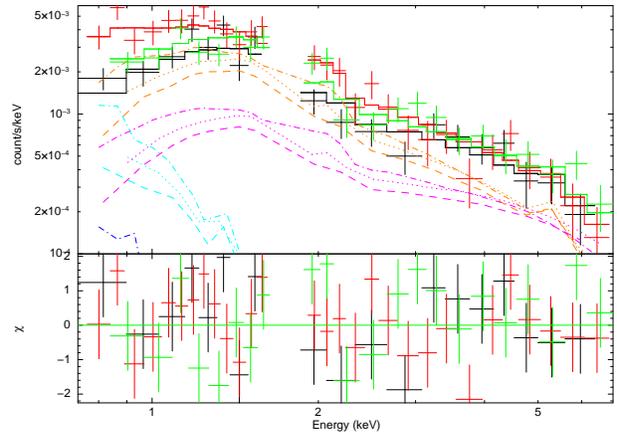}
  \end{center}
  \caption{The XIS spectra of the pre-shock (R2) region of the relic shock fitted with the model described in the text. 
         The black, red, and green crosses show the spectra of XIS0, XIS1, and XIS3, respectively.
         The total spectra of best-fit model are also plotted as solid histograms.
         The blue, light blue, magenta, and orange dotted histograms represent the LHB, MWH, CXB, and ICM components, respectively.
         The ICM component is clearly detected over the CXB.}
  \label{fig:Rpre-spec}
\end{figure}

\begin{figure}
  \begin{center}
    \includegraphics[angle=270,width=8cm]{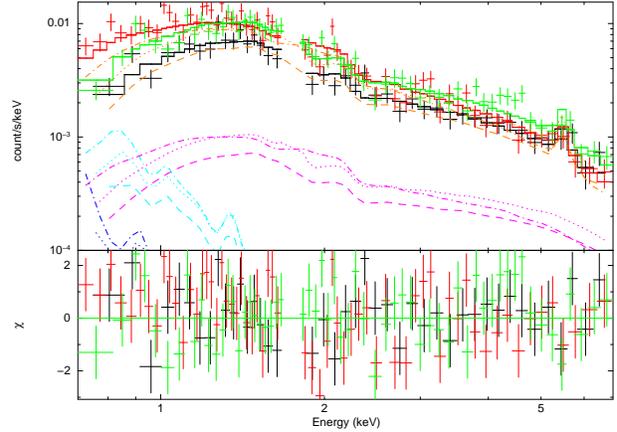}
  \end{center}
  \caption{Same as figure \ref{fig:Rpre-spec}, but for the post-shock (R3) region.}
  \label{fig:Rpost-spec}
\end{figure}

\begin{figure}
  \begin{center}
    \includegraphics[angle=270,width=8cm]{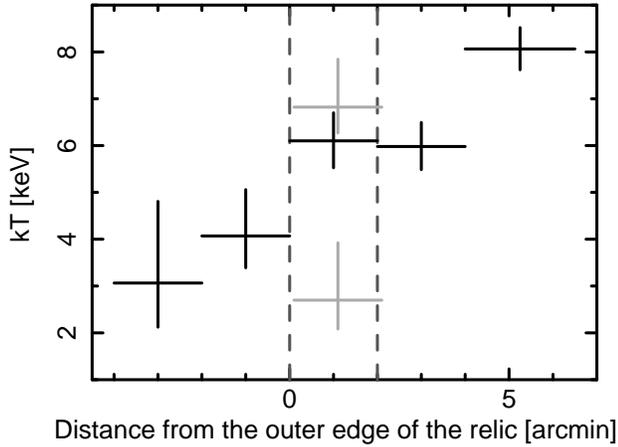}
  \end{center}
  \caption{
The temperature profile across the relic shock region. Horizontal axis represents the angular distance from the relic outer edge.
Inward and outward directions from the relic outer edge is positive and negative, respectively. Black crosses shows the results 
from region R1, R2, R3, R4, and R5. Light gray crosses show the results when the region R3 (post shock) is divided into 
R3east and R3west regions. Only statistical errors are displayed. 
The positions of the inner and outer edges of the relic are displayed by dark gray dotted lines. 
  }
  \label{fig:relic_temp}
\end{figure}

Figure \ref{fig:Rpre-spec} and \ref{fig:Rpost-spec} represent the XIS spectra of the pre-shock (R2) and post-shock (R3) regions
for the relic shock analysis, fitted with the above-mentioned model, respectively. Even in the pre-shock region where 
the surface brightness is relatively low, the ICM component is clearly detected over the CXB thanks to the low and stable 
NXB of the XIS. The fitting results are summarized in table \ref{tab:relic}, where the first, second, and third errors are statistical, 
CXB systematic, and NXB systematic, respectively. 
Even in the pre-shock region, where the ICM emission is relatively faint, the systematic errors are not much more 
than the statistical ones. The temperature profile across the relic is shown in figure \ref{fig:relic_temp}, where 
horizontal axis represents the angular distance from the relic outer edge. Inward and outward directions from the relic outer edge is 
positive and negative, respectively. Black crosses shows the results from region R1, R2, R3, R4, and R5. 
Light gray crosses show the results when the region R3 (post-shock) is divided into the east and west. 
Only statistical errors are displayed. 
The positions of the inner and outer edges of the relic are displayed by dark gray dotted lines. 
A temperature drop is surely seen at the outer edge of the relic, and we also confirmed the temperature gradient 
in the relic region along the relic's longer axis.

\begin{table*}
  \caption{Fitting results of the relic shock regions.}
  \begin{center}
    \begin{tabular}{lcccc} 
      \hline
      Region Number   & $kT$ (keV)\footnotemark[$*$]    & abundance\footnotemark[$*$]   & $\chi^2/d.o.f$ & $\Delta_{\rm CXB}$ (\%)\footnotemark[$\dagger$] \\ 
       \hline
      R1               &$3.06^{+1.74+1.05+0.57}_{-0.94-0.92-0.53}$  &$ 0.2$(fixed)   &$24.06/31$      &  38                    \\
      R2 (pre-shock)   &$4.07^{+0.99+0.69+0.40}_{-0.68-0.68-0.42}$  &$ 0.2$(fixed)   &$72.46/66$      &  33                    \\
      R3 (post-shock)  &$6.10^{+0.60+0.22+0.13}_{-0.58-0.24-0.18}$  &$ 0.2$(fixed)   &$205.76/161$    &  35                    \\
      R4               &$5.98^{+0.51+0.16+0.10}_{-0.50-0.16-0.14}$  &$0.24^{+0.07+0.00+0.00}_{-0.07-0.00-0.01}$  &$185.74/178$   & 38   \\
      R5               &$8.06^{+0.46+0.07+0.07}_{-0.44-0.07-0.09}$  &$0.16^{+0.05+0.00+0.00}_{-0.05-0.00-0.00}$  &$457.43/419$   & 32   \\
        \hline
      R3east           &$2.70^{+1.22+0.96+0.38}_{-0.62-0.77-0.21}$  &$ 0.2$(fixed)    &$17.87/17$      &  59                    \\
      R3west           &$6.82^{+1.02+0.47+0.34}_{-0.55-0.21-0.11}$  &$ 0.2$(fixed)    &$187.50/170$   & 41                      \\
        \hline
       \\
         \multicolumn{2}{@{}l@{}}{\hbox to 0pt{\parbox{180mm}{\footnotesize
       \footnotemark[$*$] The first, second, and third errors are statistical, CXB systematic, and NXB systematic, 
             respectively.
       \par \noindent
        \footnotemark[$\dagger$] CXB fluctuations at the 90 \% confidence level estimated with equation (\ref{eq:CXB}).
     }\hss}}
    \end{tabular}
    \label{tab:relic}
  \end{center}
\end{table*}

\begin{figure}
  \begin{center}
    \includegraphics[angle=270,width=8cm]{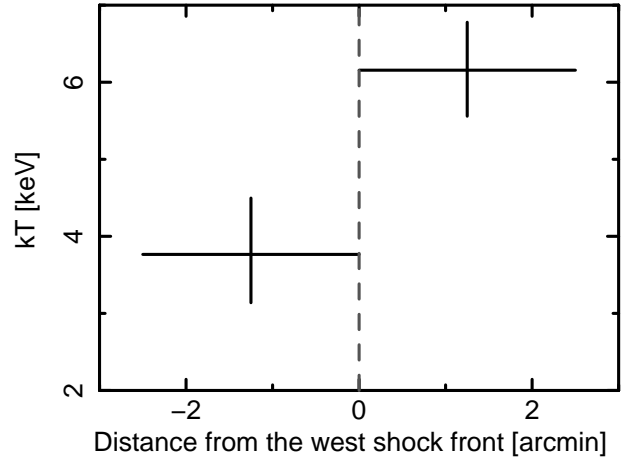}
  \end{center}
  \caption{
The temperature profile across the west shock region.  Horizontal axis represents the angular distance from the possible shock front.
Inward and outward directions from the possible shock front is positive and negative, respectively
(see cyan annuli in figure \ref{fig:region}). Only statistical errors are displayed.
}
  \label{fig:west_temp}
\end{figure}

We performed similar analysis for the west shock regions. Again, the ICM component is clearly detected over CXB 
even in the pre-shock (W1) region.
The fitting results are summarized in table \ref{tab:west}, and temperature profile is shown 
in figure \ref{fig:west_temp}. Again, we see a clear temperature drop across the expected shock front.

\begin{table*}
  \caption{Fitting results of the west shock regions.}
  \begin{center}
    \begin{tabular}{lcccc} \hline
      Region Number   & $kT$ (keV)\footnotemark[$*$]        & abundance    & $\chi^2/d.o.f$ & $\Delta_{\rm CXB}$ (\%)\footnotemark[$\dagger$] \\ \hline
     W1 (pre-shock)   & $3.76^{+0.73+0.49+0.31}_{-0.63-0.66-0.54}$  & $0.2$ (fixed)  & $89.32/83$  & 30     \\
     W2 (post-shock)  & $6.16^{+0.62+0.25+0.13}_{-0.60-0.28-0.19}$  & $0.2$ (fixed)  & $173.22/158$  & 34       \\ \hline 
       \\
         \multicolumn{2}{@{}l@{}}{\hbox to 0pt{\parbox{180mm}{\footnotesize
       \footnotemark[$*$] The first, second, and third errors are statistical, CXB systematic, and NXB systematic, 
             respectively.
       \par \noindent
        \footnotemark[$\dagger$] CXB fluctuations at the 90 \% confidence level estimated with equation (\ref{eq:CXB}).
     }\hss}}
    \end{tabular}
    \label{tab:west}
  \end{center}
\end{table*}

\subsection{Search for the Non-thermal Inverse Compton X-rays from the Relic}
It is expected that the non-thermal X-rays are emitted through the inverse Compton scattering of the CMB photons
by the same electron population attributed to the radio relic. To constrain this component, the spectra in the radio relic region
(R3 in figure \ref{fig:temp_plot}) are fitted by the following model;
\begin{eqnarray}
 const*[apec_{\rm LHB} + wabs*(apec_{\rm MWH} + powerlaw_{\rm CXB}\\ \nonumber  + apec_{\rm ICM} +powerlaw_{\rm IC})] 
\end{eqnarray}
where $powerlaw_{\rm IC}$ represents the inverse Compton component. Again, all the parameters in the background
components were fixed to the values obtained from the background field results.
We fixed $const$ to be unity for XIS1 and allowed it to vary freely for XIS0 and XIS3. The other parameters
except $const$ are common among all XIS sensors.
The temperature of the ICM component
was fixed to the value (6.10 keV) obtained in the relic shock region analysis because the expected inverse Compton component
is much weaker than the thermal ICM one. 
We fixed the photon index of the inverse Compton component to be 2.1 considering that \citet{van Weeren2012} reported that 
the integrated radio spectral index of the relic is $-1.1$. Systematic errors of the CXB and NXB
were taken into account in the same way as in the candidate shock region analysis. The fitting results are summarized in 
table \ref{tab:IC}. 
Because the lower limit of the normalization is consistent with zero (taking into account of 3 $\sigma$ errors),
we conclude that the inverse Compton component is not detected. 
Considering both the statistical and systematic errors, we derive an upper limit of the inverse Compton component
in 0.3-10 keV at the 90 \% confidence level. The resultant upper limit on the flux is $2.4 \times 10^{-13}$ erg s$^{-1}$ cm$^{-2}$
in a $2' \times 8'$ rectangular area. 
The spectral fit in the rectangular region is done by excluding point source regions. To calculate the upper limit, we assume that
the X-ray emission is uniformly distributed within this rectangular region so we can directly compare with the radio flux.
We tried to fit the data allowing the photon index of the $powerlaw_{\rm IC}$ component to vary in order to confirm the component
is really consistent with the inverse Compton interpretation. However, the fit did not work well and we obtained no useful results.

\begin{table*}
\caption{Best fit parameters of the spectrum model with inverse Compton components.}
\begin{center}
\begin{tabular}{ccc} \hline
      components         & $kT$ (keV) or $\Gamma$                & normalization\footnotemark[$\ddagger$]              \\ \hline
      $apec_{\rm ICM}$     & $6.10$ (fixed)\footnotemark[$*$]       & $3.67^{+0.64+0.37+0.15}_{-0.63-0.37-0.22} \times 10^{-2}$     \\
      $powerlaw_{\rm IC}$  & $2.1$ (fixed)\footnotemark[$\dagger$]  & $2.03^{+1.51+0.39+0.24}_{-1.51-0.39-0.19} \times 10^{-3}$      \\ \hline 
 & $\chi^2/d.o.f$ & $200.88/161$                      \\ \hline
 \\
\multicolumn{2}{@{}l@{}}{\hbox to 0pt{\parbox{180mm}{\footnotesize
\footnotemark[$*$]The value obtained from spectral fitting of the relic post shock region summarized in table \ref{tab:relic}.
\par\noindent
\footnotemark[$\dagger$] Photon index derived from the integrated radio spectral index \citep{van Weeren2012}.
\par\noindent
\footnotemark[$\ddagger$] Normalizations in the $apec$ code and powerlaw component are written in the same way 
                           as in table \ref{tab:BGD}.\\
                           The errors are represented as in table \ref{tab:relic}.
}\hss}}
\end{tabular}
\label{tab:IC}
\end{center}
\end{table*}

\section{Discussion}

\subsection{Temperature Distribution along the Collision Axis}
\begin{figure}
  \begin{center}
    \includegraphics[width=8cm,angle=0]{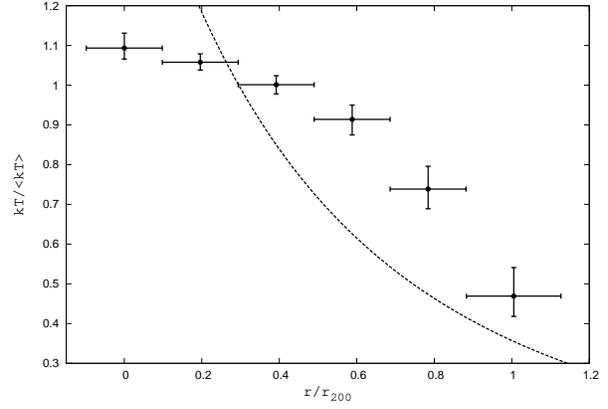}
  \end{center}
  \caption{
           Comparison of the scaled temperature profile along the collision axis toward the north with a fitting function for relaxed clusters
           proposed by \citet{Burns2010}, which is based on cosmological N-body + hydrodynamical simulation results. 
           The temperature and radius are normalized by the mean temperature and virial radius, respectively. 
           Crosses and dotted lines represent our results and the fitting function, respectively. The temperatures are higher than 
           the values of fitting function around $r \sim 0.7 r_{200}$, which might be because of the heating by the merger shock, 
           disturbed shape of the gravitational potential, and/or disturbed entropy distribution.
           }
  \label{fig:temp+burns}

\end{figure}

We compare the temperature decline in the outer part of the 1RXS J0603.3+4214 with that of typical relaxed clusters.
First, we compare our results with a scaled temperature profile proposed by \citet{Burns2010} for dynamically relaxed clusters,
which is based on cosmological N-body + hydrodynamic simulation results.
To get the scaled profile, the temperature and radius are normalized by the mean temperature and virial
radius, respectively. We adopt the mean temperature $<kT>=7.8$ keV \citep{ogrean2013}. The Virial radius is calculated
from the following equation \citep{Henry2009},
\begin{equation}
\label{eq:virial}
r_{200}=2.77\pm 0.02 h_{70}^{-1} {\rm Mpc}\frac{\left({kT}/{10 \ {\rm keV}}\right)^{1/2}}{E(z)},
\end{equation}
where $h_{70}$ is the Hubble constant normalized by the value $H_0=70$ Mpc$^{-1}$ km s$^{-1}$ and 
$E(z)=[\Omega_{0}(1+z)^3+1-\Omega_{0}]^{1/2}$.
The resultant virial radius is $r_{200}=2.21$ Mpc. We applied these values to the results presented in table \ref{tab:temp}
in the north of the main X-ray peak, which is shown in figure \ref{fig:temp+burns} by crosses.
On the other hand, the scaled temperature profile outside $0.2 r_{200}$ proposed by \citet{Burns2010} is,
\begin{equation}
\label{eq:temp}
\frac{kT}{<kT>}=A\left[1+B\frac{r}{r_{200}}\right]^{\alpha},
\end{equation}
where the best fit parameters are $A=1.74 \pm 0.03$, $B=0.64 \pm 0.10$, $\alpha=-3.2 \pm 0.4$. This fitting function is also
plotted in figure \ref{fig:temp+burns} by dotted lines.
 Considering that the 1RXS J0603 shows clear signatures of a major merger,
it is not so surprising that the observed profile does not seems to match with the fitting function. 
The obtained temperatures are higher than the values of the fitting function
around $r \sim 0.7 r_{200}$, which might be because of the heating by the merger shock, disturbed shape of the 
gravitational potential, and/or disturbed entropy distribution.

\begin{figure}
  \begin{center}
    \includegraphics[width=8cm,angle=0]{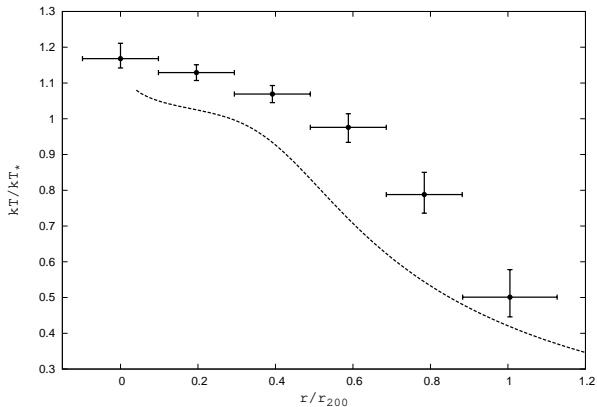}
  \end{center}
  \caption{ Comparison of the scaled temperature profile along the collision axis toward the north with a universal temperature
            profile obtained from Suzaku X-ray and Subaru weak-lensing observations \citep{Okab14}. 
           Crosses and dotted lines represent our results and the universal profile, respectively. 
           Clearly, the measured temperatures are systematically higher than the universal profile in all radii,
           though the central part and outermost data could be marginally consistent with the universal profile
           taking into account of the errors of $kT_{*}$. Note that the our data shows a convex profile whereas the universal profile is 
           concave around $r \sim 0.7 r_{200}$.
           }
  \label{fig:temp+okabe}
\end{figure}

Next, we compare our results with a universal temperature profile obtained by \citet{Okab14} from Suzaku X-ray and Subaru weak-lensing
observations of four relaxed galaxy clusters. The proposed profile is
\begin{equation}
 \frac{kT}{kT_{*}}=\biggl( \tilde{r}_0^{-1} \frac{r}{r_{200}} \biggr)^{\frac{3}{5}\alpha - \frac{2}{5}\gamma} 
                  \biggr[ 1 + ( \tilde{r}_0^{-1} \frac{r}{r_{200}} )^{\beta}  \biggl]^{-(\frac{2}{5}\delta - \frac{2}{5}\gamma + \frac{3}{5} \alpha)/\beta},
\end{equation}
where $\alpha=1.16^{+0.17}_{-0.12}$, $\beta=5.52^{+2.87}_{-2.64}$, $\gamma=1.82^{+0.28}_{-0.30}$, $\delta=2.72^{+0.34}_{-0.35}$, and 
$\tilde{r}_0 = 0.45^{+0.08}_{-0.07}$. The normalization factor of the temperature is
\begin{equation}
 kT_{*} = 1.27^{+0.24}_{-0.19} {\rm keV} \biggr[ \frac{M_{200} E(z)}{10^{14} h^{-1}_{70} M_{\odot}} \biggl]^{2/3},
\end{equation}
where $M_{200}$ is the total mass inside $r_{200}$. Though both $M_{200}$ and $r_{200}$ should be determined through direct mass measurements 
with gravitational lensing technique ideally, we do not have available lensing data for this cluster. Thus, we utilize $r_{200}$ derived from
equation (\ref{eq:virial}) and calculate $M_{200}$ simply assuming that the mean density inside $r_{200}$ is 200 times of the critical density 
of the universe. As a result, $M_{200}=1.23 \times 10^{15} M_{\odot}$ and $kT_{*} = 7.30$ keV. 
Figure \ref{fig:temp+okabe} shows a comparison of our results (crosses) with the profile of \citet{Okab14} (dotted lines).
It is clear that the measured temperatures are systematically higher than the universal profile in all radii.
This suggests that the normalization determination does not work very well, which is not so surprising considering that
we did not use gravitational lensing data. The central part and outermost data could be marginally consistent with the universal profile
taking into account of the errors of $kT_{*}$. Note that the our data shows a convex profile whereas the universal profile is concave around
$r \sim 0.7 r_{200}$.

\subsection{Shocks}
\begin{figure}
  \begin{center}
    \includegraphics[width=8cm,angle=270]{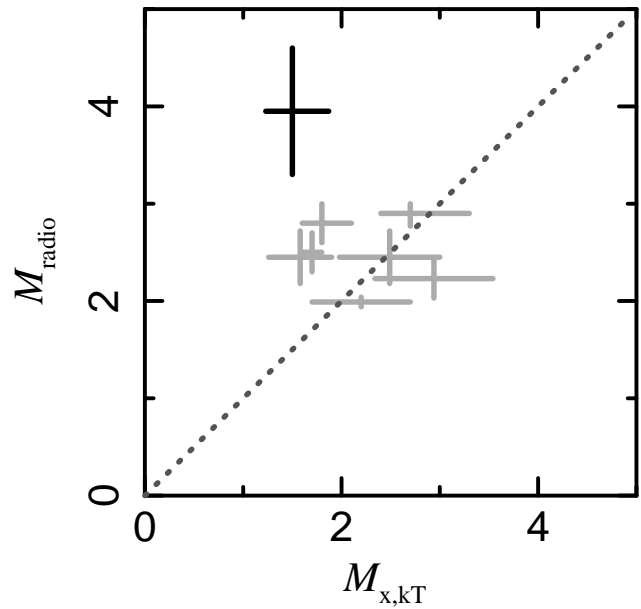}
  \end{center}
  \caption{Mach numbers derived from the radio spectral index ($M_{\rm radio}$) plotted against those from the X-ray 
           temperature measurements ($M_{X,kT}$), which is an updated version of figure 8 in \citet{A&K13}.
           The gray dotted lines represent those of $M_{\rm radio}=M_{X,kT}$.
           The error bars of $M_{X,kT}$ is only statistical and at the one sigma level. 
           The toothbrush relic (a black cross) seems to be a rather extreme example.}
  \label{fig:mach}
\end{figure}

Here, we derive the Mach number from the temperature difference of the shock candidate region using the
a Rankine-Hugoniot relation as follows,
\begin{equation}
\frac{T_2}{T_1}=\frac{5{M_X}^4+14{M_X}^2-3}{16{M_X}^2},
\label{rankine-hugoniot}
\end{equation}
where the $T_1$ and $T_2$ denote the temperature in the pre-shock and post-shock regions, respectively, and we assume
that the specific heat ratio $\gamma = 5/3$. Applying the equation (\ref{rankine-hugoniot}) to 
the region R2 and R3 of the relic shock results,
we obtain the $M_X = 1.50^{+0.37+0.25+0.14}_{-0.27-0.24-0.15}$, where the first, second, and third errors are statistical, CXB systematic, 
and NXB systematic at the 90 \% confidence level, respectively. 
This means that $M_X =1.50^{+0.28}_{-0.24}$ at the 1 $\sigma$ confidence level. On the other hand,
the radio spectral index results indicate $M_{\rm radio} = 3.3 \sim 4.6$ \citep{van Weeren2012} on the assumption of a simple DSA theory.
Thus, our results show clear discrepancy between them even including both statistical and systematic errors, 
which strongly suggests that a simple DSA theory does not hold at least for this object and that there should be other parameters 
that control the shock acceleration. 
Figure \ref{fig:mach} shows Mach numbers derived from the radio spectral index ($M_{\rm radio}$) plotted against those from the X-ray 
temperature measurements ($M_{X,kT}$), 
which is an updated version of figure 8 in \citet{A&K13}. In addition to the results of toothbrush relic
(a black cross), those of CIZA2242 north and south \citep{Stro14,Akam15}, A2256 \citep{Tras15}, Coma \citep{Thie03,Akam13}, 
A3667 north west \citep{Hind14,Akam12a} and south east \citep{Hind14,A&K13}, and A3376 \citep{Kale12,Akam12b} are displayed.
The error bars of $M_{X,kT}$ is only statistical and at the one sigma level. 
The toothbrush relic (a black cross) seems to be a rather extreme example.

A temperature gradient across the relic long axis was suggested by \citet{ogrean2013} and also confirmed in our data as shown
in figure \ref{fig:relic_temp}. 
This means that the temperature of the region R3 could be underestimated, which lead the underestimation of the
Mach number. To estimate this effect, we derived the Mach number using the region R2 and R3west temperatures. The resultant Mach number 
becomes $M_X = 1.67^{+0.43+0.27+0.16}_{-0.28-0.25-0.15}$. 
Therefore, this effect is quite limited and the discrepancy between the radio and our results still remains.
We searched for a corresponding temperature gradient in the pre-shock region, dividing R2 region into the east and west
and performing the spectral analysis as in subsection \ref{ss:an_ca_sh_re}. However, no significant temperature difference
is found because of the poor statistics.

Considering Suzaku's point spread function (PSF), whose half power diameter is $\sim 2'$, 
the temperature difference between the neighboring regions
could be somewhat underestimated. Especially, the temperature of the fainter and cooler pre-shock region would be overestimated
because of the contamination from the brighter and hotter post-shock region. To check this, we measured the temperature of 
the region $1'$ ahead of the relic outer edge.
The obtained results is $kT=2.85^{+0.93+0.76+0.43}_{-0.58-0.61-0.49}$ keV. This is $\sim$ 1.2 keV lower than the pre-shock temperature.
This temperature decline is partly because of global temperature decline trend shown in figure \ref{fig:temp_plot}, partly
because of less contamination from the post-shock region. In other word, the overestimation of pre-shock region temperature due to PSF 
should be $\sim$ 1 keV at most, which could be comparable to the statistical and CXB systematic errors at most.

If the ICM components that are not directly associated with the shock but have different temperature is accidentally located 
along the line-of-sight, we would have incorrect temperature difference.
To investigate this possibility, we fit the spectra with a two-temperature ICM model, 
but the fitting results did not improve significantly. 
At least, there is no compelling evidence to introduce two-temperature ICM model. 
Also, this unexpected temperature difference could be
a result of complicated dynamical history of the cluster. Indeed, both a peculiar linear shape of the ``toothbrush'' 
and the temperature gradient along the longer axis of the relic strongly suggest
that this cluster does not undergo a simple bimodal merger. \citet{Brug12} proposed a triple merger scenario, 
where interaction of multiple shocks generate a linear shape shock front like ``toothbrush'' relic.

In a re-acceleration scenario \citep{Brun01}, 
where the electrons in the relic have already accelerated once somewhere, the spectrum has a memory 
of the past acceleration history. In other words, the present shock Mach number alone cannot determine the relic spectrum.
If the past acceleration process is relevant to the higher Mach number shocks such as virial shocks \citep{Ryu03,Vazza09}, 
relic shock Mach number would be lower than what is expected from radio observations.
A non-linear acceleration model \citep{Malk01}, where the shock structures are modified by the interactions between
the thermal ICM and accelerated non-thermal particles, is another possibility to explain this kind of discrepancy.
The relation between the radio spectrum index and Mach number is not simple and will depend on the details of the modeling.

A similar analysis is performed also for the western shock region and the obtained Mach number becomes 
 $M_{X} =1.63^{+0.32+0.20+0.12}_{-0.28-0.27-0.22}$ at the 90 \% confidence level, which is consistent with 
the XMM-Newton results ($M=1.7^{+0.41}_{-0.42}$ at the $1 \sigma$ level confidence) based on the X-ray surface brightness analysis 
\citep{ogrean2013}. In general, the temperature measurement is less seriously affected by the line-of-sight structures. 
Note that some kind of three dimensional modeling is necessary to get the information of the density structure from 
the X-ray surface brightness distribution. On the other hand, our result could be affected by Suzaku's moderate spatial resolution
as mentioned above. Again, it is probable that a temperature difference and Mach number are underestimated to a certain extent.
Nevertheless, we obtained results consistent with \citet{ogrean2013} in an independent way.

\subsection{Constraint on the Magnetic Field Strength at the Radio Relic}
We constrain the magnetic field strength from the comparison of the radio synchrotron flux and an inverse Compton 
X-ray upper limit \citep{Rybi79}. The following procedure is basically the same as in \citet{Suga09}.
The typical energy, $h \nu'$, of photons scattered through inverse Compton processes 
by electrons with energy $\gamma m_{\rm e} c^2$ is $h \nu' \simeq 4 \gamma^2 h \nu / 3$, where $h \nu$ is the photon energy 
before scattering. With the typical CMB photon energy at $z=0.225$ ($h \nu \simeq 8.1 \times 10^{-4} {\rm eV}$), 
the range of the electron's relativistic Lorentz factor corresponding to the 0.3-10 keV inverse Compton X-rays is 
$5.3 \times 10^{2} < \gamma < 3.0 \times 10^3$. On the other hand, the synchrotron critical frequency of emitted
by electrons with magnetic field strength $B$ is $(\nu_c / {\rm MHz}) \simeq 3.3 \times 10^2 (\gamma/10^4)^2 (B / \mu \rm{G})$ 
assuming a homogeneous pitch angle distribution. Thus, the typical synchrotron frequency range emitted by the electrons 
in the energy range mentioned above is $9.1 \times 10^{-1} (B / \mu {\rm G}) < (\nu_c/{\rm MHz}) < 3.0 \times 10 (B/ \mu {\rm G})$. 
\citet{van Weeren2012} reported that the radio flux of the toothbrush relic is 319.5 mJy at 1.382 GHz and 
that the spectral index is $\alpha=-1.1$. Therefore, the radio flux corresponding to the energy range of 0.3-10 keV inverse 
Compton X-ray is $F_{\rm sync} = 2.7 \times 10^{-14} (B/ \mu {\rm G})^{-0.1}$ erg s$^{-1}$ cm$^{-2}$, 
with a monochromatic approximation for a single electron's spectrum and on the assumption of a single power-law electron 
energy distribution.

It is well-known that the flux of the inverse Compton scattering of CMB photons and synchrotron radiation 
from the same electron population has the following relation:
\begin{equation}
  \frac{F_{\rm IC}}{F_{\rm sync}}=\frac{U_{\rm CMB}}{U_{\rm mag}}=\frac{U_{\rm CMB}}{B^2/8\pi},
  \label{eq:IC-syn}
\end{equation}
where $U_{\rm CMB}$ and $U_{\rm mag}$ are the energy density of the CMB photons and magnetic field, respectively. 
With $U_{\rm CMB}=9.5 \times 10^{-13}$ erg cm$^{-3}$ at $z=0.225$, above-mentioned $F_{\rm sync}$, and obtained upper limit of $F_{\rm IC}$,
the lower limit of the magnetic field strength in the relic
becomes $B>1.6 {\rm \  \mu G}$. This is consistent with the equipartition magnetic field energy, $B_{\rm eq} = 9.2 {\rm \ \mu G}$, 
\citep{van Weeren2012}.
Note that our results are free from the assumption of energy equipartition between cosmic-ray electrons and magnetic field.

The upper limit on the inverse Compton flux might be sensitive to the temperature determination of the thermal ICM component.
We checked how the temperature changes of the ICM component affect the lower limit on the inverse Compton flux in the spectral
analysis. As a result, a 1 keV increase and decrease of the ICM temperature cause a 24 \% increase and 27 \% decrease of the
upper limit flux, respectively. These also result in an 12 \% decrease and 12 \% increase of the lower limit on magnetic field strength.
It seems odd at first glance that the temperature increase (or decrease) of the ICM causes increase (or decrease)
of the inverse Compton component. The best fit values and statistical errors of the inverse Compton component normalization are less 
sensitive to the ICM temperature change. On the other hand, the systematic errors are more sensitive and becomes larger (or smaller)
when the ICM temperature becomes higher (or lower), which is the main factor governing the resultant upper limit flux.
In any cases, magnetic fields of $\sim \mu$ G level is still inferred.

We also investigated how changes in the radio spectral index affect the results. 
The 0.1 increase and decrease of the absolute value of the radio spectral (and X-ray photon) index cause 
a 31 \% and 62 \% decrease of the upper limit flux, respectively. 
These cause a 50 \% increase and 19 \% decrease of the lower limit on the magnetic field strength, respectively.
Note that the energy range of electrons attributed to the observed synchrotron radiation is most likely higher than 
that to the inverse Compton component in our analysis. While the range of electron's Lorentz factor attributed to 
100 MHz - 1 GHz synchrotron radiation is $5.5 \times 10^3 ( B / \mu {\rm G})^{-0.5} < \gamma < 1.7 \times 10^4 (B/\mu {\rm G})^{-0.5}$,
that to the 0.3-10.0 keV inverse Compton component is $5.3 \times 10^2 < \gamma < 3.0 \times 10^3$ as discussed in the above.
Fixing the normalization in the radio observation range, therefore, increase (or decrease) of the absolute value of the 
radio spectral index causes increase (or decrease) of synchrotron flux corresponding to the inverse Compton component, 
which results in increase (or decrease) of the lower limit on the magnetic field strength.
Generally thought, the above results all indicate a magnetic field at the $\sim \mu {\rm G}$ level.
However, this might not be the case if the radio spectrum is not a single power-law because we rely on the extrapolation 
of the synchrotron spectrum. Lower frequency radio and higher energy X-ray observations are crucial in this regard. 
LOFAR and ASTRO-H will make a significant contribution to resolve this problem.

\subsection{Energy Budget of the Radio Relic Region}
The energy densities of the thermal ICM, non-thermal electrons, and magnetic field in the radio relic region can be estimated
from our results, which are basic and crucial parameters to investigate the origin and physical states of the radio relic.
For simplicity, the radio relic region is assumed to be a cylinder whose radius and height are 868 kpc and 434 kpc, respectively.
From the normalization of the apec model in the spectral analysis, the electron number density of the thermal ICM 
in the relic region is estimated to be $n_{\rm e} = 3.54 \times 10^{-4}$ cm$^{-3}$. With this value and the temperature of this region 
($kT=6.10$ keV), the energy density of the thermal ICM is $U_{\rm th}=8.6 \times 10^{-12}$ erg cm$^{-3}$, assuming that the mean molecular
weight is 0.6. With the lower limit of the magnetic field strength ($B>1.6 {\rm \  \mu G}$), the energy density of the magnetic field
is $U_{\rm mag} > 1.0 \times 10^{-13}$ erg cm$^{-3}$. As a result, $U_{\rm mag}/U_{\rm th} > 1.2 \times 10^{-2}$, 
which means that the magnetic energy could be more than a few \% of the thermal one and that the ICM evolution and 
structures could be somewhat affected by the magnetic field.
From the upper limit of the inverse Compton component, the energy density of the relativistic electrons corresponding to 
0.3-10 keV X-ray band (or, $5.3 \times 10^2 < \gamma < 3.0 \times 10^3$) becomes $U_{\rm e} < 3.6 \times 10^{-14}$ erg cm$^{-2}$.
This indicates $U_{\rm e}/U_{\rm th} < 4.3 \times 10^{-3}$, though the contribution from lower energy electrons is not included 
in this calculation, which could be dominant in the energy density of the non-thermal electron populations.

\section{Conclusions}
We observed the field around the ``toothbrush'' radio relic in the galaxy cluster 1RXS J0603.3+4214 with Suzaku. From the XIS analysis,
we found that the Mach number at the outer edge of the relic derived from the temperature difference is significantly lower than
the value derived from the radio spectral index assuming a simple DSA theory \citep{van Weeren2012}. 
This suggests that a simple DSA theory does not hold
at least for this object. The re-acceleration scenario could explain this discrepancy. Alternatively, interactions of multiple shocks
owing to complicated dynamical history would resolve both this problem and the peculiar linear shape of the relic.
We also confirmed that the Mach number of the possible shock,
which is elongated towards the west of the relic, is consistent with the value obtained from XMM-Newton data \citep{ogrean2013} 
in an independent way.
We searched for the non-thermal inverse Compton component in the relic region and set an upper limit on the flux of 
$2.4 \times 10^{-13}$ erg cm$^{-2}$ s$^{-1}$. Comparing the synchrotron radio flux with this upper limit, 
we obtained a lower limit on the magnetic field strength of $B>1.6 {\rm \ \mu G}$, 
which means that the magnetic energy could be more than a few \% of the thermal energy.

\bigskip

The authors would like to thank T. Akahori, K. Sato, S. Shibata, and H. Ohno for helpful comments. 
We are also grateful to the Suzaku operation team for their support in planning and executing this observation.
MT is supported in part by Japan Society for the Promotion of Science (JSPS) KAKENHI Grant Number 26400218. 
HA is supported by a Grant-in-Aid for JSPS Fellows (22-606). SRON is supported financially by NWO, 
the Netherlands Organization for Scientific Research.
RJW is supported by NASA through Einstein Postdoctoral grant number PF2-130104 awarded by the Chandra X-ray Center, which is operated 
by the Smithsonian Astrophysical Observatory for NASA under contract NAS8-03060.


\end{document}